\documentclass[aps,twocolumn,letterpaper,superscriptaddress,showpacs,preprintnumbers,amsmath,amssymb,prl]{revtex4}

\usepackage{ifpdf}
\usepackage{epsfig}
\usepackage{epstopdf}
\usepackage{graphicx}

\usepackage{graphicx}
\usepackage{dcolumn}
\usepackage{bm}

\begin{document}
\title{Double core hole production in N$_2$: Beating the Auger clock
}

\author{L. Fang}
\email[Corresponding author. ]{lifang@slac.stanford.edu}
\affiliation{Physics Department, Western Michigan University, Kalamazoo, MI 49008, USA}

\author{M. Hoener}
\affiliation{Physics Department, Western Michigan University, Kalamazoo, MI 49008, USA}
\affiliation{Advanced Light Source, Lawrence Berkeley National Laboratory, Berkeley, CA 94720,~USA}

\author{O. Gessner}
\affiliation{Ultrafast X-ray Science Laboratory Chemical Sciences Division Lawrence Berkeley National Laboratory, Berkeley, CA 94720,~USA}

\author{F. Tarantelli}
\affiliation{Dipartimento di Chimica, Universit\`{a}di Perugia, and ISTM-CNR, 06123 Perugia, Italy}

\author{S.T. Pratt}
\affiliation{Argonne National Laboratory, Argonne, IL 60439,~USA}

\author{O. Kornilov}
\affiliation{Ultrafast X-ray Science Laboratory Chemical Sciences Division Lawrence Berkeley National Laboratory, Berkeley, CA 94720,~USA}

\author{C. Buth}
\affiliation{The PULSE Institute for Ultrafast Energy Science, SLAC National Accelerator Laboratory, Menlo Park, CA 94025,~USA}
\affiliation{Department of Physics, Louisiana State University, Baton Rouge, LA 70803,~USA}

\author{M. G\"{u}ehr}
\affiliation{The PULSE Institute for Ultrafast Energy Science, SLAC National Accelerator Laboratory, Menlo Park, CA 94025,~USA}

\author{E.P. Kanter}
\affiliation{Argonne National Laboratory, Argonne, IL 60439,~USA}

\author{C. Bostedt}
\affiliation{Linac Coherent Light Source, SLAC National Accelerator Laboratory, Menlo Park, California 94025, USA}

\author{J.D. Bozek}
\affiliation{Linac Coherent Light Source, SLAC National Accelerator Laboratory, Menlo Park, California 94025, USA}

\author{P.H. Bucksbaum}
\affiliation{The PULSE Institute for Ultrafast Energy Science, SLAC National Accelerator Laboratory, Menlo Park, CA 94025,~USA}

\author{M. Chen}
\affiliation{Lawrence Livermore National Laboratory, Livermore, CA 94550,~USA}

\author{R. Coffee}
\affiliation{Linac Coherent Light Source, SLAC National Accelerator Laboratory, Menlo Park, California 94025, USA}

\author{J. Cryan}
\affiliation{The PULSE Institute for Ultrafast Energy Science, SLAC National Accelerator Laboratory, Menlo Park, CA 94025,~USA}

\author{M. Glownia}
\affiliation{The PULSE Institute for Ultrafast Energy Science, SLAC National Accelerator Laboratory, Menlo Park, CA 94025,~USA}

\author{E. Kukk}
\affiliation{Department of Physics and Astronomy, University of Turku 20014 Turku, Finland}

\author{S.R. Leone}
\affiliation{Ultrafast X-ray Science Laboratory Chemical Sciences Division Lawrence Berkeley National Laboratory, Berkeley, CA 94720,~USA}
\affiliation{Departments of Chemistry and Physics, University of California Berkeley, CA 94710, USA}

\author{N. Berrah}
\email[Corresponding author. ]{nora.berrah@wmich.edu}
\affiliation{Physics Department, Western Michigan University, Kalamazoo, MI 49008, USA}


\begin{abstract}

We investigate the creation of double K-shell holes in N$_2$ molecules via sequential absorption of two photons on a timescale shorter than the core-hole lifetime by using intense x-ray pulses from the Linac Coherent Light Source free electron laser. The production and decay of these states is characterized by photoelectron spectroscopy and Auger electron spectroscopy.  In molecules, two types of double core holes are expected, the first with two core holes on the same N atom, and the second with one core hole on each N atom. We report the first direct observations of the former type of core hole in a molecule, in good agreement with theory, and provide an experimental upper bound for the relative contribution of the latter type.
\end{abstract}

\maketitle

The development of intense, short pulse, x-ray free electron lasers (FELs) will allow the exploration of novel states of atoms, molecules and clusters, with potential impact on applications ranging from single-pulse imaging of biomolecules to high energy density materials~\cite{r1,r5,r7,r8,r44, r47}.  One intriguing possibility enabled by these sources is the ability to produce atoms and molecules with multiple electron vacancies in core orbitals through the sequential absorption of multiple photons on a timescale faster than Auger decay, and it has been suggested that double core holes (DCHs) could provide the basis for richer, sensitive, spectroscopies than conventional inner-shell photoelectron spectroscopy~\cite{r13,r19}. The different atomic sites in molecules introduce multiple possibilities for the DCH configurations, e.g., DCHs with both vacancies on a single site (DCHSS) and DCHs with single vacancies on two different sites (DCHTS).  For such states, the presence and location of the first hole is predicted to affect the energy to produce the second hole, as well as the decay mechanisms and fragmentation patterns of the resulting DCH states~\cite{r13}. The magnitude of the energy shifts of these states will provide unique spectral signatures as well as new information on the chemical environment of the core holes~\cite{r18,r19}.  While the intensity of conventional x-ray sources is too low to produce observable DCHs through sequential absorption processes, DCH states have been observed as a result of electron correlation in single-photon absorption. Unfortunately, these processes have a very low yield~\cite{r14,r15}, and only DCHSS states have been observed in this manner. At the intensities of the new x-ray FELs, photoabsorption should compete effectively with Auger decay~\cite{r33,r21}, allowing the production of both types of DCH states through sequential two-photon processes.  

In this Letter, we report the first experimental attempt to characterize the DCH states of a molecule, N$_2$, by sequential two-photon absorption from an x-ray FEL, the LCLS.  The production and decay of these states are characterized by using photoelectron spectroscopy and Auger-electron spectroscopy. The experimental results are interpreted with the help of \textit{ab initio} calculations~\cite{r45, r46} of the singly, doubly, and triply ionized states and their corresponding Auger spectra~\cite{r40}.  Both Green's function calculations and multi-reference configuration interaction (aug-cc-pCVQZ basis set) calculations were performed. These results will serve as a basis for understanding double- and multiple- core-hole states in more complex molecules. The experiment was conducted using the LCLS Atomic Molecular and Optical (AMO) physics instrument. We used photon energies of 1.0 keV and 1.1~keV, a pulse duration of 280 fs, and a fluence of $\sim$7$\times$10$^4$ photon$\cdot$\AA$^{-2}$ (the fluence at the beam focus is estimated to be $\sim$1$\times$10$^4$ photon$\cdot$\AA$^{-2}$ per pulse due to the photon beamline losses). The details of the apparatus and x-ray facility are described in Refs~\cite{r33, r21, r34}.

\begin{figure}[ht]
 \centerline{ \includegraphics[clip,width=1\linewidth]{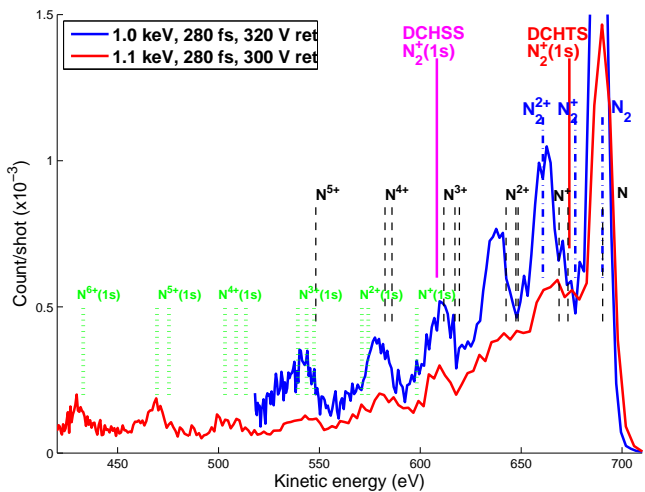}}
\vspace{-10pt}
 	\caption{(Color on line) Photoelectron spectra recorded at 1.1~keV and 1.0~keV. The electron energies in the 1 keV spectrum have been shifted by 100 eV to higher kinetic energies. Vertical markers indicate the calculated photoelectron line positions for various molecular~\cite{r40} and atomic states of N$_2$~\cite{r42}.}
  \label{fig2}
\vspace{-17pt}
\end{figure}

Figure~\ref{fig2} shows photoelectron spectra recorded at 1.0~keV and 1.1~keV photon energy and the theoretically predicted electron energies for the relevant photo-processes.  The electron energies in the 1 keV spectrum have been shifted by 100 eV to higher kinetic energies. Photoelectrons were detected in an emission direction parallel to the electric vector of the ionizing radiation. With photon energies 600 - 700 eV above the K edge of N$_2$ and pulse durations longer than the Auger-decay lifetimes ($\sim$6.4 fs for N$_2$~\cite{r23}), many ionization and decay pathways are possible, as indicated by the complex photoelectron spectrum. The labels indicate the energies of photoelectrons resulting from a number of different processes based on the calculations~\cite{r40,r42}.  These include : (1) Photoelectrons from single core-shell ionization of molecular N$_2$, N$_2^+$, and N$_2^{2+}$, where the initial state electron vacancies are located in the valence shells (dot-dashed blue lines). (2) Photoelectrons from core-shell ionization of atomic N$^{m+}$ (initial state with m valence holes) (dashed black lines). (3) Photoelectrons from core-shell ionization of atomic N$^{m+}$(1s), with a single core hole and the remaining holes in the valence shells (dotted green lines); the processes responsible for these photoelectrons result in atomic DCH production. (4) Photoelectrons from core-shell ionization of N$_2^+$(1s) with a single hole in the 1s orbital of one N atom, where the second electron is removed either from the same atom, i.e., the DCHSS process (purple solid line), or from the other N atom i.e., the DCHTS process (red solid line). Most of the spectral features are the result of multiple photon absorption by a single molecule. In particular, processes (3) and (4) present sequential two-photon inner-shell double ionization events on timescales that are fast compared to Auger decay. For the higher charge states, the reduced number of valence electrons increases the Auger lifetime, and thus increases the probability of producing atomic DCH states. The peak identification is particularly clear for N$^{4+}$(1s), N$^{5+}$(1s), and N$^{6+}$(1s) ionization, because no other photoelectrons are expected at energies below $\sim$550 eV, and the observed peaks are in good agreement with the predicted line positions~\cite{r42}. 

\begin{figure}[!ht]
 \centerline{ \includegraphics[clip,width=1\linewidth]{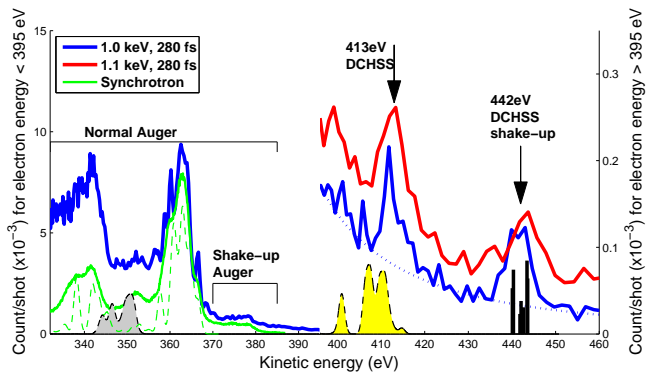}}
\vspace{-10pt}
 	\caption{(Color on line) Auger spectra from the LCLS and synchrotron experiments, and from theoretical calculations. Thick solid curves: Auger spectra recorded at 1.0 keV and 1.1 keV. The 1.1 keV spectrum is offset vertically for clarity. Solid green curve: Auger spectrum recorded with synchrotron source~\cite{r24}. Dashed curves: calculated Auger spectra (scaled) of various initial states, including SCH (green), DCHTS (grey shaded curve), DCHSS (yellow shaded curve). Black stems indicate predicted energies of DCHSS shake-up Auger electrons. Dotted curve: spline fit to background.}
\vspace{-10pt}
  \label{fig4}
\end{figure}

Figure~\ref{fig2} clearly shows a substantial peak at the expected position of the molecular DCHSS photoelectron peak. However, this energy is also close to the calculated positions of peaks associated with core-ionization of a triply valence ionized N atom, and these processes cannot be distinguished in the photoelectron data. However, unlike the photoelectron peak associated with DCHSS from neutral N$_2$, Auger decay of the molecular DCHSS state arising from neutral N$_2$ is expected to produce an Auger electron with a kinetic energy several tens of eV above the main Auger peaks.  
Fig.~\ref{fig4} shows the corresponding Auger spectra for N$_2$, where the solid curves are the experimental data and the dashed curves are our theoretical results. Auger electrons were detected in an emission direction perpendicular to the electric vector of the ionizing radiation. The Auger electron spectrum in the kinetic energy range from  $\sim$330 eV to  $\sim$450 eV is shown, along with a synchrotron-based spectrum~\cite{r24} and our calculated normal (non-resonant) N$_2$ Auger spectrum. In the region 360-370 eV, the normal Auger spectrum~\cite{r25,r26,r28}, which results from decay of the main SCH state of N$_2$, dominates. Between 370~eV and 390 eV, weak features (satellites) are observed that correspond to Auger decay from excited (shake-up) SCH states. Above $\sim$400 eV, no Auger electrons of N$_2$ have been observed with conventional x-ray sources. In contrast, the new spectra shown in Fig.~\ref{fig4} exhibit two new features lying 50-80~eV above the main SCH Auger peaks and with kinetic energies of 413~eV and 442~eV. These signals are clearly associated with Auger processes since the kinetic energies of the peaks do not vary with the photon energy. Contributions from photoelectrons observed in Fig.~\ref{fig2} are strongly suppressed in Fig.~\ref{fig4} due to the selected electron emission direction perpendicular to the light polarization. Our calculation shown in Fig.~\ref{fig4} (yellow shaded curve) predicts peaks at $\sim$410 eV, which are produced by the Auger decay of the main DCHSS state of N$_2^{2+}$ involving mainly  2$\sigma_u$ and 1$\pi_u$ electrons. In light of the expected underestimate in theory of 2-4 eV due to partially unaccounted electron relaxation in the final states~\cite{r29}, the prediction agrees very well with the observed peak position. We conclude that the newly observed Auger electron signal at $\sim$410 eV kinetic energy stems from the decay of DCHSS states and is associated with so-called Auger hypersatellites~\cite{r14,r15}. We estimate the DCHSS signal intensity to be $\sim$1\% of the main Auger peak signal between 355 eV- 370 eV ($\pm$5eV). Experiments have also been performed at the LCLS to characterize the angular distribution of these hypersatellite electrons~\cite{r22}. 

Although calculations of the Auger decay of DCHSS shakeup states have not been performed explicitly, the energy splitting between the DCHSS main Auger peak and the DCHSS shake-up Auger peaks can be estimated by the energy difference between the calculated DCHSS photoline and the DCHSS shake-up photoline. The predicted energies of the DCHSS shake-up Auger electrons are indicated by black stem lines in Fig.~\ref{fig4}. The good agreement of the calculated positions with the peak at 442 eV suggests that this peak might correspond to the Auger decay of the DCHSS shake-up states. 

\begin{figure}[!ht]
 \centerline{ \includegraphics[clip,width=0.7\linewidth]{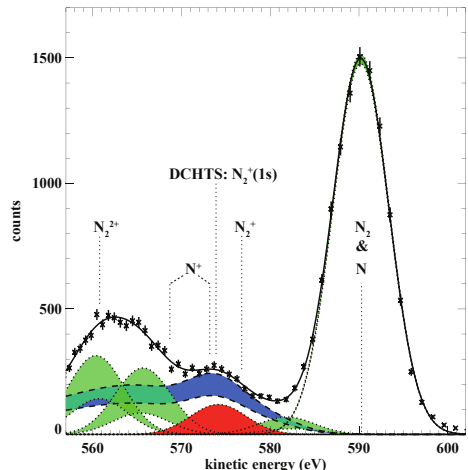}}
\vspace{-10pt}
 	\caption{(Color on line) Photoelectron spectrum (crosses with error bars) recorded at 1.0 keV photon energy and 480 V spectrometer retardation voltage. The blue, green, and red shaded areas mark the correlated uncertainties for various spectral components (see text). The constant sum of all components is marked by the solid black curve. Calculated line positions are marked by vertical dashed lines. The red shaded area allows for an upper limit estimate of DCHTS contributions.}
\vspace{-10pt}
  \label{fig5}
\end{figure}

The detection and assignment of electrons associated with the production of the uniquely molecular DCHTS states requires improved energy resolution, which is achieved by increasing the retardation voltage in the photoelectron spectrometer to 480 V, as shown in Fig.~\ref{fig5}. The spectrum is analyzed by taking into account the contributions from single-photon shake-up/off (SUO) processes as previously determined by Svensson et al.~\cite{r30} and Kaneyasu et al.~\cite{r31}, and convoluting them with the experimental energy resolution of the current study. The dashed lines and blue shaded area in Fig.~\ref{fig5} indicate the shape and the range of uncertainty of the SUO contributions, respectively, based on the uncertainties in assigning exact peak areas in the spectrum of Svensson et al.~\cite{r30}. Note that in single photon ionization, SUO processes would generate the only contributions to the photoelectron spectrum in the kinetic energy range between $\sim$400 eV and $\sim$580 eV. Therefore, all photoelectron signal shown in Fig.~\ref{fig5} that is not contained in the main photoline at 590 eV or the SUO contributions is due to multiple photon processes that are enabled by the high intensity of the LCLS. To disentangle the major contributions generated by multiple photon ionization, a nonlinear least-squares fit is performed based on five spectral components with identical peak shapes and independent amplitudes and center energies (green and red shaded areas). The common peak shape is derived by a description of the well-defined main photoline by a Gaussian function, which gives a good approximation of the light-source- and apparatus-induced line broadening effects.

The range of possible SUO intensities results in a corresponding uncertainty range for the intensities of all other spectral contributions, as indicated by the green and red shaded areas, the borders of which indicate the upper and lower limits. However, all peak center energies and the overall fit quality are virtually unaffected by the SUO uncertainty.  Within the model of identical, apparatus-limited peak shapes, the fit result is globally unique as determined by a Monte-Carlo based sampling of the available fit starting parameter space by $\sim$10$^5$ independent fit cycles. We therefore base our spectral analysis on the comparison of the fit results with calculated line positions, as marked in Fig.~\ref{fig5} by vertical dashed lines. The theoretical peak positions mark the kinetic energies expected from 1s inner-shell ionization of neutral and valence-ionized atoms and molecules as indicated. Also marked is the expected position of the DCHTS photoelectron line that results from inner-shell ionization of an N$_2$(1s$^{-1}$) molecular core-hole state. We note that subtleties such as dynamic line shifts during molecular fragmentation are not included in the fit model. However, these are considered minor effects with respect to the conclusions that are derived here.

Taking into account the finite precision of the calculations ($\sim$1-2 eV) and the finite energy resolution of the measurements, the agreement of the experimentally derived peak positions with the theoretical expectations is mostly very good. Independent, physically motivated cross-checks, such as the analysis of the ratio of valence- to inner-shell ionization probabilities and the rate of molecular dissociation compared to the pulse length, confirm that the measured peak intensities for inner-shell ionization of N$_2^{2+}$ and N$^+$ agree qualitatively with the picture of sequential multiple photon ionization by the LCLS at the peak intensity of the experiments. The weak contribution at $\sim$582 eV cannot readily be assigned to a calculated line. Slight deviations of the mainline peak shape from a perfect Gaussian, as can be seen at the high energy side of the main photoline, may account for some part of this feature. Some signal from N$_2^+$ ionization may also contribute to this part of the photoelectron spectrum.

One spectral component is consistently found exactly at the position of the expected DCHTS photoline (574~eV). Unfortunately, this spectral region is also marked by the strongest SUO contributions. Additionally, the 1s$^{-1}$ photoline from inner-shell ionization of excited N$^+$(1s$^2$2s$^1$2p$^3$) fragments coincides with the DCHTS photoline within the theoretical and experimental uncertainties. As indicated by the red area, the estimate of the total multiple photon ionization signal at 574~eV varies between zero and 8\% of the main photoline intensity. Given the uncertainty in the contributions from overlapping spectral components, we can provide only an upper bound for the DCHTS signal rather than an absolute value. The upper limit of the DCHTS contribution is mainly defined by the minimum contribution of the N$^+$(1s$^2$2s$^1$2p$^3$) ionization signal. The lower bound of this signal is estimated by the peak at 566~eV, which is mainly attributed to inner shell ionization of N$^+$(1s$^2$2s$^2$2p$^2$), and by the branching ratio between ground and excited states of N$^+$ fragments after inner shell ionization of N$_2$~\cite{r32}. This analysis leads to an upper bound for DCHTS contributions in the photoelectron spectrum of 4\% relative to the intensity of the N$_2$(1s$^{-1}$) main photoline. This value is in qualitative agreement with the $\sim$1\% DCHSS signal described above and an estimated ratio of $\sim$1.65 between DCHTS and DCHSS signal intensities. An independent estimate of the intensity of the DCHTS process by examination of the Auger spectrum is not possible at this stage. The nature of these states marked by a single core hole per atom leads to Auger electron energies that are embedded in the SCH Auger spectrum, as indicated by the grey shaded structure near 350 eV kinetic energy in Fig.~\ref{fig4}. Furthermore, Auger decay of sequential multiple photoionization products, in addition to correlation satellites and the decay of SUO states, may also contribute to this range of the Auger electron spectrum, making it difficult to isolate the DCHTS Auger processes at this time. Note that in Fig.~\ref{fig4} only the high-energy, most intense, bands of the computed DCHTS and DCHSS Auger spectra are shown, involving ionization of the outer-valence electrons. The low-energy band of the DCHTS spectrum is calculated to fall outside the experimental range, while the DCHSS one falls in the 380-390 eV shake-up region where it is not appreciably detected by the experiment. The theoretical determination of the relative probabilities for SCH and DCH formation requires complex calculations that we hope to stimulate with the present work.  Thus, no comparison between experiment and theory has been made on this point. 

In conclusion, we have presented the first experimental results on the characterization of DCH states produced by sequential two-photon absorption in molecules.  The observation of DCHSS states is supported by photoelectron and Auger-electron spectra, as well as by theoretical calculations.  In contrast, the observation of the uniquely molecular DCHTS states remains ambiguous, and only an upper bound to their production can be determined.  Higher intensity (afforded by improved beam transport optics), improved stability of photon energy, higher electron energy resolution, and detailed studies of the intensity dependence of the electron spectra will thus be required to fulfill the potential of DCH spectroscopies. 

This work was supported by the U.S. Department of Energy, Office of Science, Basic Energy Sciences, Division of Chemical Sciences, Geosciences, and Biosciences. We thank the LCLS staff for their assistance, and A.~ Kivim\"{a}ki, T.~Jahnke and  R.~D\"{o}rner for providing us with unpublished data. We thank  Robin Santra for fruitful discussion.  We thank B. Kr\"{a}ssig for his eTOF models. FT thanks italian FIRB and PRIN grants.CBu was supported by NSF grants. MH thanks  the Alexander von Humboldt foundation for the Feodor Lynen fellowship. The LCLS is funded by DOE-BES.

\vspace{-15pt}
\bibliographystyle{apsrev}
\bibliography{DCH_PR_etof_ref}

\begin{thebibliography}{28}
\expandafter\ifx\csname natexlab\endcsname\relax\def\natexlab#1{#1}\fi
\expandafter\ifx\csname bibnamefont\endcsname\relax
  \def\bibnamefont#1{#1}\fi
\expandafter\ifx\csname bibfnamefont\endcsname\relax
  \def\bibfnamefont#1{#1}\fi
\expandafter\ifx\csname citenamefont\endcsname\relax
  \def\citenamefont#1{#1}\fi
\expandafter\ifx\csname url\endcsname\relax
  \def\url#1{\texttt{#1}}\fi
\expandafter\ifx\csname urlprefix\endcsname\relax\def\urlprefix{URL }\fi
\providecommand{\bibinfo}[2]{#2}
\providecommand{\eprint}[2][]{\url{#2}}

\bibitem[{\citenamefont{{R.~Neutze \textit{et al.}}}(2000)}]{r1}
\bibinfo{author}{\bibnamefont{{R.~Neutze \textit{et al.}}}},
  \bibinfo{journal}{Nature} \textbf{\bibinfo{volume}{406}},
  \bibinfo{pages}{752} (\bibinfo{year}{2000}).

\bibitem[{\citenamefont{{H.~N.~Chapman \textit{et al.}}}(2006)}]{r5}
\bibinfo{author}{\bibnamefont{{H.~N.~Chapman \textit{et al.}}}},
  \bibinfo{journal}{Nat.~Phys.} \textbf{\bibinfo{volume}{2}},
  \bibinfo{pages}{839} (\bibinfo{year}{2006}).

\bibitem[{\citenamefont{{J.~D.~Lindl \textit{et al.}}}(2004)}]{r7}
\bibinfo{author}{\bibnamefont{{J.~D.~Lindl \textit{et al.}}}},
  \bibinfo{journal}{Phys. Plasmas} \textbf{\bibinfo{volume}{11}},
  \bibinfo{pages}{339} (\bibinfo{year}{2004}).

\bibitem[{\citenamefont{Dyer et~al.}(2008)\citenamefont{Dyer, Bernstein, Cho,
  and Osterholz}}]{r8}
\bibinfo{author}{\bibfnamefont{G.}~\bibnamefont{Dyer}},
  \bibinfo{author}{\bibfnamefont{A.}~\bibnamefont{Bernstein}},
  \bibinfo{author}{\bibfnamefont{B.}~\bibnamefont{Cho}}, \bibnamefont{and}
  \bibinfo{author}{\bibfnamefont{J.}~\bibnamefont{Osterholz}},
  \bibinfo{journal}{Phys. Plasmas} \textbf{\bibinfo{volume}{101}},
  \bibinfo{pages}{015002} (\bibinfo{year}{2008}).

\bibitem[{r44()}]{r44}
\bibinfo{note}{{N.~Berrah \textit{et al.}}, J.~Mod.~Opt. {\bf 57}, xx (2010)
  (in press).}

\bibitem[{\citenamefont{{C. Bostedt \textit{et al.}}}(2008)}]{r47}
\bibinfo{author}{\bibnamefont{{C. Bostedt \textit{et al.}}}},
  \bibinfo{journal}{Phys.~Rev.~Lett.} \textbf{\bibinfo{volume}{100}},
  \bibinfo{pages}{133401} (\bibinfo{year}{2008}).

\bibitem[{\citenamefont{Cederbaum et~al.}(1986)\citenamefont{Cederbaum,
  Tarantelli, and Sgamellotti}}]{r13}
\bibinfo{author}{\bibfnamefont{L.~S.} \bibnamefont{Cederbaum}},
  \bibinfo{author}{\bibfnamefont{F.}~\bibnamefont{Tarantelli}},
  \bibnamefont{and}
  \bibinfo{author}{\bibfnamefont{A.}~\bibnamefont{Sgamellotti}},
  \bibinfo{journal}{J.~Chem.~Phys} \textbf{\bibinfo{volume}{85}},
  \bibinfo{pages}{6513} (\bibinfo{year}{1986}).

\bibitem[{\citenamefont{Santra et~al.}(2009)\citenamefont{Santra, Kryzhevoi,
  and Cederbaum}}]{r19}
\bibinfo{author}{\bibfnamefont{R.}~\bibnamefont{Santra}},
  \bibinfo{author}{\bibfnamefont{N.~V.} \bibnamefont{Kryzhevoi}},
  \bibnamefont{and} \bibinfo{author}{\bibfnamefont{L.~S.}
  \bibnamefont{Cederbaum}}, \bibinfo{journal}{Phys.~Rev.~Lett.}
  \textbf{\bibinfo{volume}{103}}, \bibinfo{pages}{013002}
  (\bibinfo{year}{2009}).

\bibitem[{\citenamefont{{M.~Tashiro \textit{et al}}}(2010)}]{r18}
\bibinfo{author}{\bibnamefont{{M.~Tashiro \textit{et al}}}},
  \bibinfo{journal}{J.~Chem.~Phys} \textbf{\bibinfo{volume}{132}},
  \bibinfo{pages}{184302} (\bibinfo{year}{2010}).

\bibitem[{\citenamefont{Kanter et~al.}(1999)\citenamefont{Kanter, Dunford,
  {B.~Kr\"{a}ssig}, and Southworth}}]{r14}
\bibinfo{author}{\bibfnamefont{E.~P.} \bibnamefont{Kanter}},
  \bibinfo{author}{\bibfnamefont{R.~W.} \bibnamefont{Dunford}},
  \bibinfo{author}{\bibnamefont{{B.~Kr\"{a}ssig}}}, \bibnamefont{and}
  \bibinfo{author}{\bibfnamefont{S.~H.} \bibnamefont{Southworth}},
  \bibinfo{journal}{Phys.~Rev.~Lett.} \textbf{\bibinfo{volume}{83}},
  \bibinfo{pages}{508} (\bibinfo{year}{1999}).

\bibitem[{\citenamefont{{S.~H.~Southworth \textit{et al}}}(2003)}]{r15}
\bibinfo{author}{\bibnamefont{{S.~H.~Southworth \textit{et al}}}},
  \bibinfo{journal}{Phys.~Rev.~A} \textbf{\bibinfo{volume}{67}},
  \bibinfo{pages}{062712} (\bibinfo{year}{2003}).

\bibitem[{\citenamefont{{L.~Young \textit{et al.}}}(2010)}]{r33}
\bibinfo{author}{\bibnamefont{{L.~Young \textit{et al.}}}},
  \bibinfo{journal}{Nature} \textbf{\bibinfo{volume}{466}}, \bibinfo{pages}{56}
  (\bibinfo{year}{2010}).

\bibitem[{\citenamefont{{M.~Hoener \textit{et al.}}}(2010)}]{r21}
\bibinfo{author}{\bibnamefont{{M.~Hoener \textit{et al.}}}},
  \bibinfo{journal}{Phys.~Rev.~Lett.} \textbf{\bibinfo{volume}{104}},
  \bibinfo{pages}{253002} (\bibinfo{year}{2010}).

\bibitem[{r45()}]{r45}
\bibinfo{note}{J. Schirmer and A. Barth, Z. Phys. A {\bf 317}, 267 (1984). A.
  Tarantelli and L.S. Cederbaum, Phys. Rev. A {\bf 46}, 81 (1992). F.
  Tarantelli, Chem. Phys. {\bf 329}, 11 (2006)}.

\bibitem[{r46()}]{r46}
\bibinfo{note}{T. H. Dunning, J. Chem. Phys. {\bf{90}}, 1007 (1989). D. E.
  Woon, T. H. Dunning, J. Chem. Phys. {\bf 103}, 4572 (1995).}

\bibitem[{\citenamefont{Tarantelli et~al.}(1991)\citenamefont{Tarantelli,
  Sgamellotti, and Cederbaum}}]{r40}
\bibinfo{author}{\bibfnamefont{F.}~\bibnamefont{Tarantelli}},
  \bibinfo{author}{\bibfnamefont{A.}~\bibnamefont{Sgamellotti}},
  \bibnamefont{and} \bibinfo{author}{\bibfnamefont{L.~S.}
  \bibnamefont{Cederbaum}}, \bibinfo{journal}{J.~Chem.~Phys}
  \textbf{\bibinfo{volume}{94}}, \bibinfo{pages}{523} (\bibinfo{year}{1991}).

\bibitem[{\citenamefont{Bozek}(2009)}]{r34}
\bibinfo{author}{\bibfnamefont{J.~D.} \bibnamefont{Bozek}},
  \bibinfo{journal}{Eur.~Phys.~J.~Special Topics}
  \textbf{\bibinfo{volume}{169}}, \bibinfo{pages}{129} (\bibinfo{year}{2009}).

\bibitem[{\citenamefont{Chen et~al.}(1979)\citenamefont{Chen, Laiman,
  Crasemann, Aoyagi, and Mark}}]{r42}
\bibinfo{author}{\bibfnamefont{M.~H.} \bibnamefont{Chen}},
  \bibinfo{author}{\bibfnamefont{E.}~\bibnamefont{Laiman}},
  \bibinfo{author}{\bibfnamefont{B.}~\bibnamefont{Crasemann}},
  \bibinfo{author}{\bibfnamefont{M.}~\bibnamefont{Aoyagi}}, \bibnamefont{and}
  \bibinfo{author}{\bibfnamefont{H.}~\bibnamefont{Mark}},
  \bibinfo{journal}{Phys.~Rev.~A} \textbf{\bibinfo{volume}{19}},
  \bibinfo{pages}{2253} (\bibinfo{year}{1979}).

\bibitem[{\citenamefont{{B.~Kempgens \textit{et al.}}}(1996)}]{r23}
\bibinfo{author}{\bibnamefont{{B.~Kempgens \textit{et al.}}}},
  \bibinfo{journal}{J.~Phys.~B} \textbf{\bibinfo{volume}{29}},
  \bibinfo{pages}{5389} (\bibinfo{year}{1996}).

\bibitem[{r24()}]{r24}
\bibinfo{note}{A.~Kivim\"{a}ki (unpublished data from Elletra synchrotron light
  source)}.

\bibitem[{\citenamefont{\r{A}gren}(1981)}]{r25}
\bibinfo{author}{\bibfnamefont{H.}~\bibnamefont{\r{A}gren}},
  \bibinfo{journal}{J.~Chem.~Phys} \textbf{\bibinfo{volume}{75}},
  \bibinfo{pages}{1267} (\bibinfo{year}{1981}).

\bibitem[{\citenamefont{{W.~E.~Moddeman \textit{et al.}}}(1971)}]{r26}
\bibinfo{author}{\bibnamefont{{W.~E.~Moddeman \textit{et al.}}}},
  \bibinfo{journal}{J.~Phys.~B} \textbf{\bibinfo{volume}{55}},
  \bibinfo{pages}{2317} (\bibinfo{year}{1971}).

\bibitem[{\citenamefont{Liegener}(1983)}]{r28}
\bibinfo{author}{\bibfnamefont{C.}~\bibnamefont{Liegener}},
  \bibinfo{journal}{J.~Phys.~B} \textbf{\bibinfo{volume}{16}},
  \bibinfo{pages}{4281} (\bibinfo{year}{1983}).

\bibitem[{\citenamefont{Schulte et~al.}(1996)\citenamefont{Schulte, Cederbaum,
  and Tarantelli}}]{r29}
\bibinfo{author}{\bibfnamefont{H.~D.} \bibnamefont{Schulte}},
  \bibinfo{author}{\bibfnamefont{L.~S.} \bibnamefont{Cederbaum}},
  \bibnamefont{and}
  \bibinfo{author}{\bibfnamefont{F.}~\bibnamefont{Tarantelli}},
  \bibinfo{journal}{J.~Phys.~B} \textbf{\bibinfo{volume}{105}},
  \bibinfo{pages}{11108} (\bibinfo{year}{1996}).

\bibitem[{r22()}]{r22}
\bibinfo{note}{{J.~Cryan \textit{et al.} (to be published).}}

\bibitem[{\citenamefont{{S.~Svensson \textit{et al.}}}(1992)}]{r30}
\bibinfo{author}{\bibnamefont{{S.~Svensson \textit{et al.}}}},
  \bibinfo{journal}{J.~Phys.~B} \textbf{\bibinfo{volume}{25}},
  \bibinfo{pages}{135} (\bibinfo{year}{1992}).

\bibitem[{\citenamefont{{T.~Kaneyasu \textit{et al.}}}(2008)}]{r31}
\bibinfo{author}{\bibnamefont{{T.~Kaneyasu \textit{et al.}}}},
  \bibinfo{journal}{J.~Phys.~B} \textbf{\bibinfo{volume}{41}},
  \bibinfo{pages}{135101} (\bibinfo{year}{2008}).

\bibitem[{r32()}]{r32}
\bibinfo{note}{T.~Jahnke, PhD Thesis (2005) (unpublished)}.

\end{thebibliography}

\end{document}